\documentclass[preprint,showpacs,pre]{revtex4}%
\usepackage{amsfonts}
\usepackage{amsmath}
\usepackage{amssymb}
\usepackage{graphicx}%
\begin{document}
\preprint{ }
\title[ ]{On the law of increase of entropy for nonequililbrium systems}
\author{A. P\'{e}rez-Madrid}
\affiliation{Departament de F\'{\i}sica Fonamental , Facultat de F\'{\i}sica, Universitat
de Barcelona. Diagonal 647, 08028 Barcelona, Spain}
\author{}

\begin{abstract}
Under the assumption of a smooth full phase-space distribution function we
prove that the nonequilibrium entropy $S$ which is considered as a functional
of the distribution vector for an N-body system possesses a lower bound and
therefore can not decrease. We also compute the rate of change of $S$ ,
$\partial S/\partial t$, showing that this is non-negative and having a global
minimum at equilibrium. As an aplication we obtain a generalization of the
(Bhatnager-Gross-Krook) BGK relaxation model.

\end{abstract}
\maketitle

\section{Introduction}

As it is known, the \ second law of Thermodynamics has been based on Kelvin's
and Claussius's principles which establish that work can not be produced
without spending energy. These principles are based on experiments but to my
knowledge no theoretical derivation based on first principles has been
given\cite{cohen}. For an adiabatically isolated system the second law
establishes that the entropy can not decrease and is maximum at
equilibrium\cite{landau}.

To provide a theoretical proof of the second law or the law of increase of
entropy for an isolated system first it is necessary to properly define the
nonequilibrium entropy which would coincide with the thermodynamic entropy at equilibrium.

The equilibrium entropy is a well established concept. According to Boltzmann
the entropy of a microstate $X$ of a macroscopic system is a number,
$S_{B}(X)=k_{B}\log\left\vert \Gamma_{M(X)}\right\vert $, with $\left\vert
\Gamma_{M}\right\vert $\ being the volume of the region of the phase space
$\Gamma_{M}$\ corresponding to the macrostate $M=M(X)$ and $k_{B}$ is the
Boltzmann constant. The macrostate $M$ is all of a group of states $Y$ such
that $M(Y)=M(X)=M$. For Boltzmann, every adiabatically isolated system in its
natural evolution moves towards an state of maximum disorder -the state of
equilibrium, which corresponds with the greatest possible volume in phase
space. Gibbs, on the other hand, defines entropy as a functional of the
distribution function $F$ in phase space, $S_{G}=-k_{B}\int F\ln FdX$, and
both entropies coincide at equilibrium. Out of equilibrium the situation is
more ambiguous, due to the fact that the Gibbs entropy is a constant of motion
under the microscopic dynamic given through the Liouville equation, it has
been thought that macroscopic processes require coarse-graining. This point
was worked out first by Gibbs and P. Ehrenfest \& T. Ehrenfest. On the other
hand, some authors argue that the relevant entropy for understanding
thermodynamic irreversibility is the Boltzmann entropy and not the Gibbs
entropy \cite{lebowitz}-\cite{goldstein}.

In this scenario, unlike treatments based on the definition of coarse-grained
entropies\cite{kampen}, here we propose a generalization of the fine-grained
Gibbs entropy postulate in terms of the set of reduced distribution functions
of the system, the distribution vector, as the nonequilibrium entropy of an
N-body isolated system\cite{agusti}, \cite{agusti2}. Under the assumption of
smooth phase-space densities we prove that this nonequilibrium entropy can not
decrease and is not a constant of motion under the dynamic given through the
BBGKY \ (Bogolyubov-Born-Green-Kirkwood-Yvon) hierarchy of
equations\cite{balescu}. Here, we assume smoothness to exclude multifractal
phase-space distributions which some simulations have shown\cite{agusti}. 

It is a fact that the Liouville equation describes an incompressible flow of
representative points in phase space, however this is not the case with
respect to the BBGKY\ hierarchy. Additionally, unlike an incompressible flow a
compressible flow produces entropy, thus in this fact resides the reason of
the behavior of the entropy we propose . Moreover, this entropy coincides with
the Thermodynamic entropy at equilibrium. Hence, we are able to compute the
entropy production for an isolated system, which is non-negative (zero at
equilibrium), therefore conferring the character of the nonequilibrium
potential of the system to this entropy production. Is this a theoretical
proof of the second law of Thermodynamics? it depends to the credit one gives
to the nonequilibrium entropy we propose, nonetheless this is the closest
definition to the microscopic dynamics of the isolated system we know.

We have structured the paper as follows. In section 2, we introduce the
Hamiltonian dynamics of the N-body system obtaining the generalized Liouville
equation. In section 3, we analyze the properties of the nonequilibrium
entropy, compute the entropy production and obtain the kinetic equation for
the one-particle reduced distribution function. Finally in section 4, we
emphasize our main conclusions.

\section{Hamiltonian dynamics}

We will consider a N-body system whose Hamiltonian results from the addition
of a kinetic energy term and a term coming from the interaction between the
particles
\begin{equation}
H=\sum_{j=1}^{N}\frac{\mathbf{p}_{j}^{2}}{2m}+\frac{1}{2}\sum_{j\neq k=1}%
^{N}\phi\left(  \left\vert \mathbf{q}_{j}-\mathbf{q}_{k}\right\vert \right)
\text{ \ ,} \label{hamiltonian}%
\end{equation}
with $m$ being the mass of a particle, and $\phi\left(  \left\vert
\mathbf{q}_{j}-\mathbf{q}_{k}\right\vert \right)  \equiv\phi_{jk}$ the
interaction potential. Moreover, the equations of motion are
\begin{equation}
\overset{\cdot}{\mathbf{q}}_{i}=\frac{\partial H}{\partial\mathbf{p}_{i}%
}\text{ \ , }\overset{\cdot}{\mathbf{p}}_{i}=-\frac{\partial H}{\partial
\mathbf{q}_{i}}\text{ \ .} \label{eq_motion}%
\end{equation}
The statistical description of the system can be performed in terms of the
full phase-space distribution function $F(x^{N},t)$, where $x^{N}%
=\{x_{1},.....,x_{N}\}$ and $x_{j}=(\mathbf{q}_{j},\mathbf{p}_{j})$ or
alternatively in terms of the distribution vector $\mathbf{f}$ \cite{balescu},
both completely equivalent but \ the last more appropriated for nonequilibrium
systems in order to show irreversibility. Here,%
\begin{equation}
\mathbf{f}\equiv\left\{  f_{o},f_{1}(x_{1},t),f_{2}(x^{2},t),...,f_{N}%
(x^{N},t)\right\}  \text{ \ } \label{distri_vec}%
\end{equation}
is the set of all the s-particle reduced distribution functions,
$s=0,.....,N$, where the s-particle reduced distribution functions
\begin{equation}
f_{s}=\frac{N!}{(N-s)!}\int F(x^{N},t)\text{ }dx_{s+1}...dx_{N}\text{ ,}
\label{reduceddistribution}%
\end{equation}
are obtained by integrating over $N-s$ particles with $f_{o}=1$. The dynamics
of the reduced distribution functions can by obtained from the Liouville
equation%
\begin{equation}
\frac{\partial}{\partial t}F=[H,F]_{p}\text{ \ ,} \label{liouville}%
\end{equation}
where $[..,..]_{p}$ is the Poisson bracket, by using Eq.
(\ref{reduceddistribution}). One obtains the generalized Liouville
equation\cite{agusti}-\cite{balescu}%
\begin{equation}
\frac{\partial}{\partial t}\mathbf{f}(t)=\mathcal{L}\mathbf{f}(t)\text{ \ ,}
\label{gene_liou}%
\end{equation}
which is a compact way of writing the BBGKY hierarchy of equations. In this
equation, $\mathcal{L}$ is the generalized Liouvillian whose diagonal part
$\mathcal{PL}$ is defined through\cite{agusti2}%

\begin{equation}
\langle s\left\vert \mathcal{PL}\right\vert s^{\prime}\rangle=\delta
_{s,s^{\prime}}\left\{  \sum_{j=1}^{s}L_{j}^{o}+\sum_{j<n=1}^{s}%
L_{j,n}^{\prime}\right\}  \text{ \ ,} \label{diagonalliouville_1}%
\end{equation}
where $\left\vert s\right\rangle $ represent the s-particle state defined
through $\left\langle s\right\vert \mathbf{f\mid}x^{s}\rangle=f_{s}(x^{s})$,
with $x^{s}=\{x_{1},.....,x_{s}\}$, $s=0,.....,N$. The s-particle states have
the property $\langle s\mid s^{\prime}\rangle=\delta_{s,s^{\prime}}$ such that
$\sum_{s=0}^{N}\mid s\rangle\langle s\mid=1$ constitutes the closure relation.
Moreover, $L_{j}^{o}=\left[  H_{j}^{o},...\right]  _{P}$ ,where $H_{j}%
^{o}=\mathbf{p}_{j}^{2}\diagup2m$, and $L_{j,n}^{\prime}=\left[
H_{j,n}^{\prime},...\right]  _{P}$ , with $H_{j,n}^{\prime}=\frac{1}{2}%
\phi_{j,n}$. In addition, the non-diagonal part $\mathcal{QL}$ is given
by\cite{agusti2}
\begin{equation}
\langle s\left\vert \mathcal{QL}\right\vert s^{\prime}\rangle=\delta
_{s^{\prime},s+1}\int\left\{  \sum_{j=1}^{s}L_{j,s+1}^{\prime}\right\}
dx_{s+1}\text{ .} \label{nondiagonal}%
\end{equation}
Finally, in view of Eqs. (\ref{diagonalliouville_1}) and (\ref{nondiagonal})
and using the properties of the s-particle states $\left\vert s\right\rangle
$, by multiplying by $\left\vert x^{s}\right\rangle $ to the right and by
$\left\langle s\right\vert $ to the left of (\ref{gene_liou}) one obtains one
of the components of the hierarchy
\begin{equation}
\frac{\partial}{\partial t}f_{s}-\sum_{j=1}^{s}L_{j}^{o}f_{s}-\sum_{j<n=1}%
^{s}L_{j,n}^{\prime}f_{s}=\sum_{j=1}^{s}\int L_{j,s+1}^{\prime}f_{s+1}%
dx_{s+1}\text{ ,} \label{onecomponent}%
\end{equation}
where the right hand side is the collision term coupling $f_{s}$ with
$f_{s+1}$. In the case of $s=1$ Eq. (\ref{onecomponent}) reduces to%
\begin{equation}
\frac{\partial}{\partial t}f_{1}+\mathbf{p}_{1}\frac{\partial}{\partial
\mathbf{q}_{1}}f_{1}=-\int\mathbf{F}_{1,2}\frac{\partial}{\partial
\mathbf{p}_{1}}f_{2}\text{ }dx_{2}\text{ \ ,} \label{onecomponent_1}%
\end{equation}
where $\mathbf{F}_{1,2}=-\nabla_{1}\phi_{1,2}$\cite{harris}.

\section{Entropy production and irreversibility}

Our statistical theory is based on the definition of the nonequilibrium
entropy proposed in \cite{agusti}, \cite{agusti2}%
\begin{align}
S  &  =-k_{B}\text{tr}\left\{  \mathbf{f}\ln\left(  \mathbf{f}_{eq}%
^{-1}\mathbf{f}\right)  \right\}  +S_{eq}\nonumber\\
&  =-k_{B}\sum_{s=1}^{N}\frac{1}{s!}\int f_{s}\ln\frac{f_{s}}{f_{eq,s}%
}\;dx_{1}.....dx_{s}\text{ }+S_{eq}\text{ } \label{gibbs}%
\end{align}
which generalizes the Gibbs entropy postulate, where $S_{eq}$ is the
equilibrium entropy coinciding with the thermodynamic entropy and
$\mathbf{f}_{eq}$ is the equilibrium distribution vector satisfying
$\mathcal{L}\mathbf{f}_{eq}=0$. The nonequilibrium entropy (\ref{gibbs})
reaches its maximum value $S=S_{eq}$ at the equilibrium state, when
$\mathbf{f=f}_{eq}$.

At this point, it should be clarified that a description of the system in
terms of the full phase-space N-particle distribution function is justified
for equilibrium systems for which the representative ensemble is distributed
uniformly through the phase space. Nonetheless, the nonuniformity of
nonequilibrium systems leads to a random clusterization caused by the
interaction between different s-particle clusters, and the same global
quantity may be obtained for\ an infinite number of realizations. This fact is
taken into account in the BBGKY hierarchy\cite{martynov} which explicitly
incorporates the interaction between different s-particle clusters through the
nondiagonal part $\mathcal{QL}$ of the generalized Liouvillian $\mathcal{L}$.
The nonequilibrium system behaves as a random mixture of s-particle systems
and, therefore the total entropy should be the sum of all the s-particle
entropies. The randomness creates entropy as we will see later on.

More interestingly here is to establish the direction of change of this
entropy in a natural process. To discern this we first note that since
$f_{s}\ln$ $f_{s}/f_{eq,s}$ is a convex function\cite{grad}
\begin{equation}
f_{s}\ln\frac{f_{s}}{f_{eq,s}}\geqq f_{s}-f_{eq,s}\text{ \ ( for }f_{s}%
\geqq0\text{ \ and }f_{eq,s}>0\text{)} \label{convexity}%
\end{equation}
and due to the fact that $f_{s}$ and $f_{eq,s}$ both are smooth phase-space
densities in the s-particle phase space%
\begin{equation}
\int f_{s}\ln\frac{f_{s}}{f_{eq,s}}\;dx_{1}.....dx_{s}\text{ }\geqq0\text{ .}
\label{positivity}%
\end{equation}
Thus, given that $f_{s}$ is bounded and since $f_{eq,s}>0$ there must exist a
constant $C$ such that
\begin{equation}
\int f_{s}\ln\frac{f_{s}}{f_{eq,s}}\;dx_{1}.....dx_{s}\leqq C\text{ \ ,}
\label{bound}%
\end{equation}
for $s=1,....,N$. \ Hence, a lower bound for the nonequilibrium entropy
exists
\begin{equation}
0>S-S_{eq}\geqq-k_{B}\sum_{s=1}^{N}\frac{1}{s!}C\text{ \ ,}
\label{entropy_bound}%
\end{equation}
which means that the nonequilibrium entropy (\ref{gibbs}) can not decrease.
Consequently, the law of increase of entropy (\textit{i.e. }the second law of
thermodynamics) follows, constituting this result one of the goals of the
present contribution.

The rate of change of the nonequilibrium entropy~\ (\ref{gibbs}) or entropy
production can be computed using Eqs. (\ref{gene_liou}), \ giving%
\begin{gather}
\frac{\partial S}{\partial t}=-k_{B}\text{tr}\left\{  \frac{\partial
\mathbf{f}}{\partial t}\ln\left(  \mathbf{f}_{eq}^{-1}\mathbf{f}\right)
\right\}  =\nonumber\\
-k_{B}\text{tr}\left\{  \mathcal{L}\mathbf{f}\ln\left(  \mathbf{f}_{eq}%
^{-1}\mathbf{f}\right)  \right\}  \text{ \ \ ,} \label{rate}%
\end{gather}
which according to Eq. (\ref{entropy_bound}) should be non-negative. In a more
explicit way, after using Eq. (\ref{onecomponent}), Eq. (\ref{rate}) can be
written as
\begin{equation}
\frac{\partial S}{\partial t}=-\frac{1}{T}\sum_{s=1}^{N}\frac{1}{s!}\sum
_{j=1}^{s}\int f_{s}\mathbf{p}_{j}\left(  -k_{B}T\frac{\partial}%
{\partial\mathbf{q}_{j}}\ln f_{eq,s}+\sum_{j\neq i=1}^{s}\mathbf{F}%
_{j,i}+\mathcal{F}_{j}\right)  dx_{1}.....dx_{s}\text{ \ }
\label{entropy_produc}%
\end{equation}
where $\mathcal{F}_{j}(x^{s})$ is$\mathcal{\ }$defined through $f_{s}%
(x^{s})\mathcal{F}_{j}(x^{s})=%
{\displaystyle\int}
\mathbf{F}_{j,s+1}f_{s+1}dx_{s+1}$, $T$ is the equilibrium kinetic temperature
taking into account that the dependence of $f_{eq,s}$ in the velocities is
given through a local Maxwellian and $\mathbf{F}_{j,i}=-\nabla_{j}\phi_{j,i}$.

The entropy production given through Eq. (\ref{entropy_produc}) vanishes at
equilibrium when $f_{s}$ coincides with $f_{eq,s}$. Moreover, because
$\mathbf{p}_{j}$ is arbitrary
\begin{equation}
\sum_{j\neq i=1}^{s}\mathbf{F}_{j,i}+\mathcal{F}_{j}^{eq}=k_{B}T\frac
{\partial}{\partial\mathbf{q}_{j}}\ln f_{eq,s}\text{ \ } \label{equiv}%
\end{equation}
is sufficient to satisfy the extremum condition $\delta\dot{S}/\delta
f_{s}\mid_{eq}=0$, with $\dot{S}\equiv\partial S/\partial t$. Here, the right
hand side of \ Eq. (\ref{equiv}) is the mean force \cite{hill} and the left
hand side is the sum of two parts, the first is the force due to the van der
Vaals interactions with the $s-1$ fixed particles different from the $j$-$th$
particle in the $s$-$th$ cluster (responsible for the compressions and
dilatations of the $s$-$th$ cluster), while the second is the average force on
particle $j$-$th$ by the remaining $N-s$ particles of the system. It is known
that Eq. (\ref{equiv}) gives rise to a hierarchy of equations the
Yvon-Born-Green (YBG) hierarchy\cite{balescu}. \ Now, by making use of the
extremum condition \ Eq.(\ref{equiv}), we can rewrite the entropy production
Eq. (\ref{entropy_produc}) as%
\begin{align}
\frac{\partial S}{\partial t}  &  =-\frac{1}{T}\frac{1}{s!}\sum_{j=1}^{N}\int
f_{s}\mathbf{p}_{j}\left(  \mathcal{F}_{j}-\mathcal{F}_{j}^{eq}\right)
dx_{1}.....dx_{s}=\nonumber\\
&  -\frac{1}{T}\sum_{s=1}^{N}\frac{1}{s!}\sum_{j=1}^{s}\int\left(
\bigtriangleup f_{s}+f_{eq,s}\right)  \mathbf{p}_{j}\left(  \mathcal{F}%
_{j}-\mathcal{F}_{j}^{eq}\right)  dx_{1}.....dx_{s}\text{ \ , \ }
\label{entropy_produc_4}%
\end{align}
where $\bigtriangleup f_{s}=f_{s}-f_{eq,s}$. The right hand side of the last
equal sign in Eq. (\ref{entropy_produc_4}) has two parts, the second of which
involving $f_{eq,n}$, provided $\mathcal{F}_{j}$ has even parity under time
reversal, vanishes. It must be pointed out that $\mathcal{F}_{j}$ has even
parity if $\ f_{s+1}(x^{s+1})=f_{1}(x_{s+1})f_{s}(x^{s})$ which is a
reasonable assumption in large enough time scales. Henceforth, after defining
the nonequilibrium current $\mathbf{J}_{j}\equiv\bigtriangleup f_{s}%
\mathbf{p}_{j}$ which vanishes at equilibrium, Eq. (\ref{entropy_produc_4})
can be rewritten
\begin{equation}
\frac{\partial S}{\partial t}=-\frac{1}{T}\sum_{s=1}^{N}\frac{1}{s!}\sum
_{j=1}^{s}\int\mathbf{J}_{j}\left(  \mathcal{F}_{j}-\mathcal{F}_{j}%
^{eq}\right)  dx_{1}.....dx_{s}\text{ \ .} \label{entropy_produc_2}%
\end{equation}
In the previous equation (\ref{entropy_produc_2}) the term under the integral
possesses the form of the product of a thermodynamic current $\mathbf{J}_{j}$
times the corresponding conjugated thermodynamic force $\left(  \mathcal{F}%
_{j}-\mathcal{F}_{j}^{eq}\right)  $, which constitutes the typical expression
of the entropy production in Nonequilibrium Thermodynamics. At this point,
following the standards of Nonequilibrium Thermodynamics\cite{degroot}, from
Eq. (\ref{entropy_produc_2}) provided the existence of an equilibrium state is
assumed, it follows
\begin{equation}
\mathbf{J}_{i}=-\sum_{j=1}^{s}\frac{\mathbf{L}_{ij}}{T}\left(  \mathcal{F}%
_{j}-\mathcal{F}_{j}^{eq}\right)  \text{ ,} \label{current}%
\end{equation}
where $\mathbf{L}_{ij}$ is a phenomenological coefficient which should be
non-negative and having even parity or in terms of the related mobility
$\mathbf{M}_{ij}=\mathbf{L}_{ij}/Tf_{s}$%
\begin{equation}
\mathbf{J}_{i}=-\sum_{j=1}^{s}f_{s}\mathbf{M}_{ij}\left(  \mathcal{F}%
_{j}-\mathcal{F}_{j}^{eq}\right)  \text{ .} \label{current_2}%
\end{equation}
The phenomenological relations (\ref{current}) or (\ref{current_2}) manifest
the fact that a restoring current arises when the system deviates from
equilibrium by the action of a thermodynamic force, this constituting a
generalization of the Onsager's regression hypotesis\cite{onsager}.
Additionally, $\mathbf{M}_{ij}$ might be a function of $f_{s}$ and $\left\{
\mathcal{F}_{j}\right\}  $, $\mathbf{M}_{ij}(f_{s},\left\{  \mathcal{F}%
_{j}\right\}  )$ in general. Therefore, after substituting Eq.
(\ref{current_2}) into Eq. (\ref{entropy_produc_2})%
\begin{equation}
\frac{\partial S}{\partial t}=\frac{1}{T}\sum_{s=1}^{N}\frac{1}{s!}%
\sum_{i,j=1}^{s}\int f_{s}\mathbf{M}_{ij}\left(  \mathcal{F}_{j}%
-\mathcal{F}_{j}^{eq}\right)  ^{2}dx_{1}.....dx_{s}\text{ }\geqq0\text{ .\ }
\label{entropy_produc_3}%
\end{equation}

To test our previous analysis, near equilibrium we can define an effective
mobility $\mathbf{M}_{ij}^{eq}=\int\mathbf{M}_{ij}f_{eq,s}dx_{1}.....dx_{s}$,
thus%
\begin{equation}
\mathbf{J}_{i}=-\sum_{j=1}^{s}f_{s}\mathbf{M}_{ij}^{eq}\left(  \mathcal{F}%
_{j}-\mathcal{F}_{j}^{eq}\right)  \label{current_4}%
\end{equation}
which by introducing the inverse mobility matrix $\mathbf{\zeta}_{ij}^{eq}$
($\sum_{j=1}^{s}\mathbf{\zeta}_{ij}^{eq}\mathbf{M}_{jl}^{eq}=\delta_{il}$),
the friction matrix, can be rewritten as%
\begin{gather}
f_{s}\left(  \mathcal{F}_{i}-\mathcal{F}_{i}^{eq}\right)  =-\sum_{j=1}%
^{s}\mathbf{\zeta}_{ij}^{eq}\mathbf{J}_{j}=\nonumber\\
-\sum_{j=1}^{s}\mathbf{\zeta}_{ij}^{eq}\mathbf{p}_{j}\left(  f_{s}%
-f_{eq,s}\right)  \text{ .} \label{force}%
\end{gather}
For $s=1$, the previous equation allows us to derive from Eq.
(\ref{onecomponent_1}) the following kinetic equation for $f_{1}\equiv f$%
\begin{equation}
\frac{\partial}{\partial t}f+\mathbf{p}\frac{\partial}{\partial\mathbf{q}%
}f+k_{B}T\left(  \frac{\partial}{\partial\mathbf{q}}\ln f_{eq}\right)
\frac{\partial}{\partial\mathbf{p}}f=\zeta^{eq}\frac{\partial}{\partial
\mathbf{p}}\mathbf{p}\left(  f-f_{eq}\right)  \label{bgk}%
\end{equation}
constituting a generalization of the BGK \cite{bhatnager} relaxation model
containing the mean force and an extra acceleration term involving
$\partial/\partial\mathbf{p}\left(  f-f_{eq}\right)  $ which distinguishes
this equation with respect to the original BGK.

\section{Conclusions}

In the framework of the BBGKY description we have found a way of generalizing
the Gibbs fine-grained entropy for an isolated N-body system leading to
irreversibility. We have proved that the nonequilibrium entropy of an isolated
N-body system defined as a convex functional of the distribution vector can
not decrease and that the rate of change of this entropy is a non-negative
quantity having a global minimum at equilibrium. As an application of those
results, after computing the entropy production, following the methods of
\ Nonequilibrium Thermodynamics we introduce a positively defined mobility
matrix which relates the nonequilibrium current with the corresponding
thermodynamic force enabling us to write the entropy production as a
positively defined quadratic form. As a consequence of introducing the
phenomenological law, we have obtained a generalization of the BGK relaxation
model which contains the mean force and an extra acceleration term involving
$\partial/\partial\mathbf{p}\left(  f-f_{eq}\right)  $. At this point it
should be mentioned that the nonequilibrium thermodynamics analysis may be
generalized to the phase space as done in the Mesoscopic Nonequilibrium
Thermodynamics (MNET) \cite{rubi}.

It should be stressed that as we have shown a proper description of the
nonequilibrium system might be given in terms of the distribution vector which
evolves according to the genreralized Liouville equation. The compressibility
inherent to the BBGKY hierarchy leads to the randomness, so that the system
can be thought either as a random mixture of s-particle systems or an
interacting mixture of compressible fluids. Hence, this description brings
about the correct macroscopic dynamics.

Due to the generality of our theory of irreversibility even the Boltzmann
irreversible equation and hence the Boltzmann H-theorem are embedded in here,
as was shown in a previous paper\cite{agusti2}. Moreover, a quantum
formulation is possible, which will be shown elsewhere.

\end{document}